\documentclass[conference]{IEEEtran}
\IEEEoverridecommandlockouts
\usepackage{cite}
\usepackage{array}

\usepackage{hyperref}
\hypersetup{colorlinks=true,linkcolor=blue,anchorcolor=blue,citecolor=blue,urlcolor=blue}

\usepackage{amsmath,amssymb,amsfonts}
\usepackage{algorithmic}
\usepackage{graphicx}
\usepackage{textcomp}
\usepackage{multirow}
\usepackage{xcolor}

\def\BibTeX{{\rm B\kern-.05em{\sc i\kern-.025em b}\kern-.08em
    T\kern-.1667em\lower.7ex\hbox{E}\kern-.125emX}}
\begin{document}

\title{{A Hybrid Graph Neural Network for Enhanced EEG-Based Depression Detection}\\
\thanks{* Corresponding author}
}

\author{\IEEEauthorblockN{1\textsuperscript{st} Yiye Wang}
\IEEEauthorblockA{\textit{School of Biological Science} \\
\textit{and Medical Engineering} \\
\textit{Southeast University}\\
Nanjing, China \\
yiyewang@seu.edu.cn}

\and
\IEEEauthorblockN{2\textsuperscript{nd} Wenming Zheng\textsuperscript{*}}
\IEEEauthorblockA{\textit{School of Biological Science} \\
\textit{and Medical Engineering} \\
\textit{Southeast University}\\
Nanjing, China \\
wenming\_zheng@seu.edu.cn}

\and
\IEEEauthorblockN{3\textsuperscript{rd} Yang Li}
\IEEEauthorblockA{\textit{School of Biological Science} \\
\textit{and Medical Engineering} \\
\textit{Southeast University}\\
Nanjing, China \\
li-yang@seu.edu.cn}

\and
\IEEEauthorblockN{4\textsuperscript{th} Hao Yang}
\IEEEauthorblockA{\textit{School of Biological Science} \\
\textit{and Medical Engineering} \\
\textit{Southeast University}\\
Nanjing, China \\
hao\_yang@seu.edu.cn}

}

\maketitle

\begin{abstract}
Graph neural networks (GNNs) are gaining increasing popularity for EEG-based depression detection. However, previous GNN-based methods inadequately consider the characteristics of depression, which limit their performance. First, neuroscience studies indicate that patients with depression exhibit both common and individualized brain abnormalities. Previous GNN-based approaches typically focus either on common graph connections to capture common brain abnormalities or on individualized connections to capture individualized patterns, which is insufficient for depression detection. Second, brain network exhibits a hierarchical structure, ranging from  channel-level graphs to  region-level graphs. This hierarchical structure varies across individuals and contains significant information relevant to detecting depression. However, previous GNN-based methods overlook this individualized hierarchical information. To address these issues, we propose a Hybrid GNN (HybGNN) that combines a Common Graph Neural Network (CGNN) branch using common connections and an Individualized Graph Neural Network (IGNN) branch employing individualized connections. The two branches capture common and individualized depression patterns, respectively, complementing each other. Furthermore, we enhance the HybGNN with a Cross-Branch Hierarchical Information Extractor (CB-HIE) to extract more task-relevant individualized hierarchical information. Extensive experiments on the MODMA and HUSM datasets demonstrate that the proposed HybGNN achieves state-of-the-art performance.


\end{abstract}


\section{Introduction}
Depression is a medical condition characterized by abnormalities of mood, cognition and neurovegetative functions, which has become a major public health concern\cite{fava2000major}. Clinically, doctors often diagnose depression through face-to-face interviews and questionnaires\cite{cummins2017generalized}. The former primarily relies on the doctor's expertise and personal judgment, while the latter depends on the patient's self-reported feelings, which can be easily concealed. Therefore, a reliable and accurate method for objective depression detection is urgently required\cite{chen2021urinary}.

Electroencephalography (EEG) provides an objective measurement of brain activity, which presents advantages over traditional methods for diagnosing depression\cite{acharya2015computer}. EEG channels are arranged on irregular grids, and traditional convolutional neural networks (CNNs)\cite{2021deprnet} struggle to effectively represent this non-Euclidean structured data. Recently, graph neural networks (GNNs), with the strength of modeling non-Euclidean structured data, have been successfully applied to the classification of EEG signals and the detection of depression\cite{2021bibm_gcn,24SGPSL,62SDGCN,hubin2022gicn,hubin2023explainableGCN,luo2023GNN}. GNN-based methods can effectively uncover disease-related patterns
and detect depression according to the variable interactions
between different EEG channels\cite{graph_survey,eeg_survey_acm}. However, these GNN-based methods are not specifically tailored for the depression detection and fail to sufficiently consider its characteristics, leading to suboptimal performance.

First, graph connections in previous GNN-based methods typically fall into two categories: common connections\cite{2021bibm_gcn,24SGPSL,62SDGCN,hubin2022gicn} and individualized connections\cite{hubin2023explainableGCN,luo2023GNN}. Common connections refer to connections that are identical across all instances, while individualized connections are adaptive to each instance. The common graph connections can
uncover the common patterns among depressed patients, ensuring stable depression detection. In contrast, the individualized connections consider individual differences and capture individualized patterns. Despite their widespread use, researchers typically select only one type of connection for the construction of graph. However, neuroscience research reveals that depressed individuals exhibit both common and individualized abnormal patterns in their brain networks\cite{2015fcdepr}. Specifically, depressed patients commonly exhibit abnormal activity in the prefrontal lobe (DMN)\cite{2015fcdepr}, while various subtypes of depression display individualized abnormal patterns in cognitive control network (CCN)\cite{subtypes}. This physiological behavior indicates that  constructing only one type of graph connection is inadequate for the complete interpretation of depression, which motivates us to integrate both types of graph connections in depression detection. 

Second, the hierarchical structure serves as a key characteristic for detecting depression, which previous GNN-based methods often overlook. This structure includes the arrangement from channel-level graphs to region-level graphs, with nodes representing channels and regions respectively\cite{region1}. The hierarchical structure varies among individuals and contains significant depression-related information\cite{region2}. Previous study investigates the hierarchical structure based on individualized brain connectivity, finding that EEG channels in the frontal lobe of depression patients tend to be partitioned into different regions in the region-level graph, whereas those in healthy individuals tend to cluster within the same region in the region-level graph\cite{hubin2017MST}.
Inspired by these studies, we aim to enhance the GNN-based model for depression detection by incorporating a module designed to extract hierarchical information from the informative individualized graph connections. However, the individualized graph connections include pseudo-connections, which introduce noise and irrelevant information\cite{Individual_noise,graph_survey}, potentially obscuring the extraction of hierarchical information. Therefore, it is crucial to mitigate the impact of such noise and task-irrelevant information during extraction process.

In this paper, we propose a Hybrid GNN (HybGNN) model that combines two graph construction methods to improve depression detection. The HybGNN consists of two branches: a Common Graph Neural Network (CGNN) branch, which constructs a common graph using common graph connections, and an Individualized Graph Neural Network (IGNN) branch, which builds individualized graphs by adaptively generating graph connections based on input instances. These two branches complement each other: the CGNN captures common depression-related patterns for robust detection, while the IGNN addresses individual differences that CGNN overlooks. Additionally, when individualized patterns are obscured by noise, the CGNN provides a reliable baseline for depression detection. 
Moreover, to equip our model with task-relevant hierarchical information, we embed a Cross-Branch Hierarchical Information Extractor (CB-HIE) into HybGNN. In CB-HIE, Gated Adaptive Graph Pooling (GAGP) is proposed, which partitions channel-level graph into region-level graph. The GAGP is performed based on the combination of informative IGNN features and robust CGNN features. In GAGP, CGNN features act as gating inputs, and through the gating mechanism, CGNN features regulate the selection of information in IGNN features, effectively controlling the node partitioning and reducing the impact of pseudo-connections and noise. Furthermore, we conduct extensive experiments on two publicly available datasets, MODMA and HUSM. The results of these experiments validate the effectiveness of our model.

\section{Related work}
\subsection{EEG-based Depression Detection}
Many studies have advanced automatic depression detection using EEG. Early approaches primarily relied on traditional machine learning with hand-crafted features. Sun et al.\cite{handcrafted2020_hubin} extracted multiple types of EEG features and evaluated the popular classifiers. Shen et al.\cite{handcrafted2023_hubin} proposed an adaptive channel fusion method to effectively optimize the spatial information. However, these methods heavily depend on labor-intensive preprocessing and feature selection, requiring specialized expertise.

Owing to the powerful feature representation capabilities of deep learning, recent methods extract task-oriented features directly from raw data and perform detection in an end-to-end manner. Seal et al.\cite{2021deprnet} proposed DeprNet, which utilizes CNN and considers both spatial information and temporal information. Additionally, GNN-based methods have been successfully applied to enhance depression detection. Sun et al.\cite{sun79CN-GCN} proposed CNGCN based on phase-locking value (PLV). Chen et al.\cite{24SGPSL} proposed SGP-SL based on predefined local similarity and global connections. However, due to the complexity of depression and brain activity, predefined graph connections are insufficient for effective EEG feature extraction. To tackle this problem, Zhu et al. \cite{hubin2022gicn} and Cui et al. \cite{62SDGCN} used learnable weight matrices as graph connections. Although these learnable graph connections yield promising results, they are common across all individuals, overlooking individual differences. Additionally, they fail to construct a region-level graph and effectively leverage hierarchical information, which limits the models' performance.

\subsection{Graph Neural Network for EEG Classification} \label{gnnec}

Although not yet fully explored in EEG-based depression recognition, existing methods in other EEG classification tasks have considered approaches for handling individual differences and extracting hierarchical information. To address individual differences in EEG-based sleep stage classification, GraphSleepNet\cite{graphsleepnet} introduced an adaptive approach that generates adaptive graph connections based on each instance. To extract hierarchical information, researchers partition channel-level nodes into regions, construct region-level graph and capture inter-regional dependencies for EEG-based emotion recognition. For instance, Ding et al.\cite{ding2023LGGNet} proposed LGGNet and established a hierarchy by manually dividing channels into regions. Given the limitations of manual division, Xue et al.\cite{xue2022ahgnn} proposed AHGNN which utilizes an adaptive graph pooling method to partition channels to regions based on Diffpool\cite{diffpool}.

Our HybGNN differs from the aforementioned methods in two key aspects: 1) To capture the common and individualized abnormal patterns of depressive patients, we simultaneously adopt the common graph connections like \cite{dgcnn} and build individualized graph connections based on attention mechanism \cite{vaswani2017attention}. 2) To better capture individual-specific hierarchical information, HybGNN leverages the individualized graph connections for channel-level nodes partition, whereas DiffPool\cite{diffpool} relies on binary graph connections and AHGNN\cite{xue2022ahgnn} employs common ones. The individualized graph connections are more informative but noisier compared to binary or common graph connections. To efficiently utilize individualized graph connections for the channel partition and region-level graph construction, we propose the GAGP module, which effectively mitigates the impact of noise and enhances the quality of the extracted hierarchical information.


\section{Methodology}

\subsection{Overview}
\begin{figure*}
\centering
\includegraphics[width=0.8\linewidth]{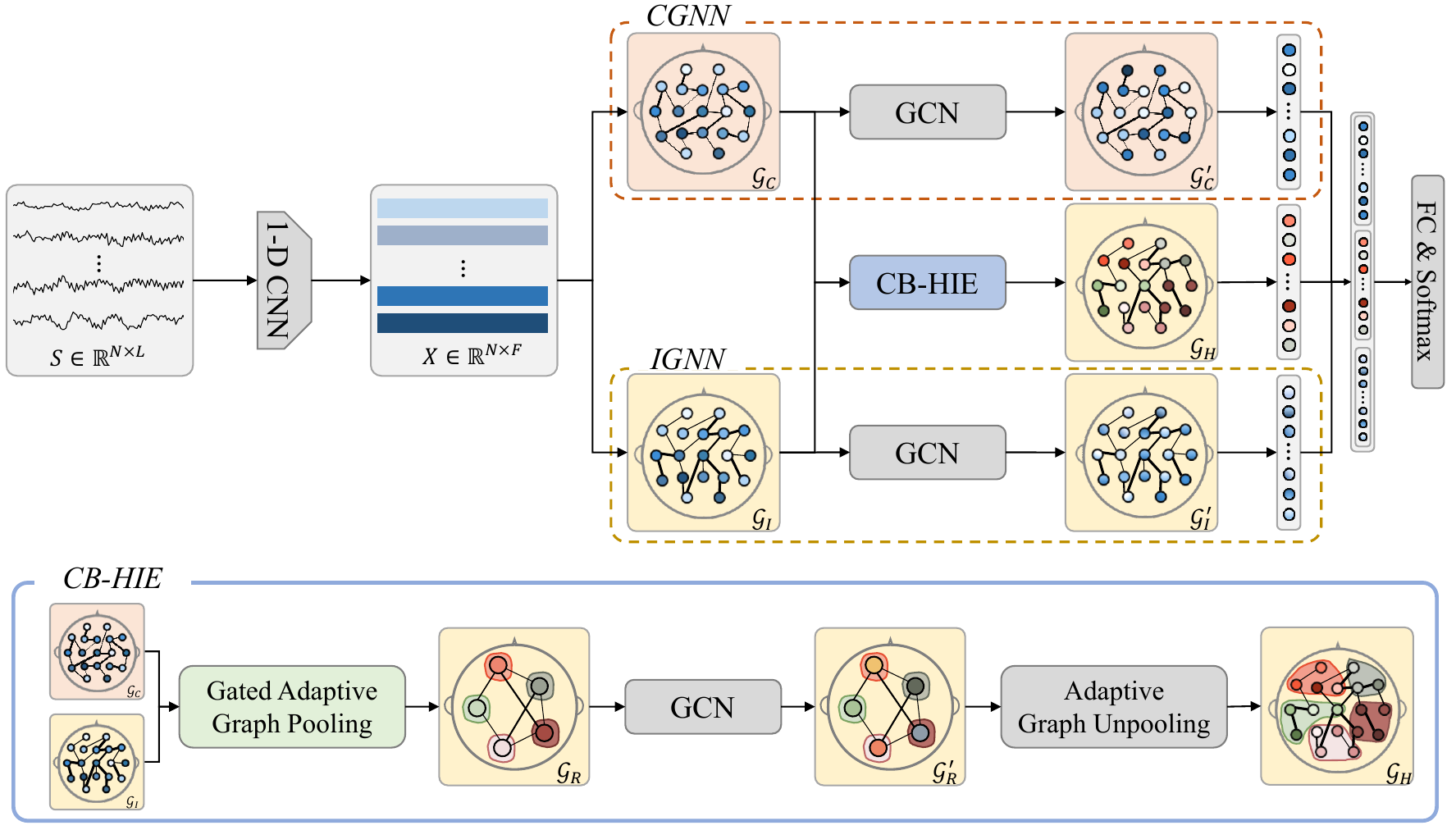} 
\caption{The framework of the proposed HybGNN. The graphs highlighted in orange use the common adjacency matrix, while those highlighted in yellow use the individualized adjacency matrix.}
\label{fig1}
\end{figure*}

The framework of the proposed HybGNN is shown in Fig.~\ref{fig1}. Two key parts in the proposed
HybGNN model are summarized as follows: 1) To identify common depression-related patterns while addressing individual differences, we propose two branches of GNNs: the CGNN (depicted in the upper branch of the figure) and the IGNN (depicted in the lower branch of the figure). 2) To further extract meaningful individualized hierarchical information, the CB-HIE is integrated into the HybGNN. Details of the model are described in the following sections.

\subsection{EEG Temporal Feature Extraction} \label{2.2}
We propose a 1-D CNN module to extract temporal dependencies within each channel, operating along the time dimension while preserving spatial information, to facilitate graph construction in the next step.

A raw EEG sample is represented as $\boldsymbol{S}\in {\mathbb{R}}^{N\times {L}}$, where $N$ denotes the number of EEG channels and each channel has $L$ time stamps. The 1-D CNN is employed to extract temporal features from the raw EEG sample $\boldsymbol{S}$. The output of the 1-D CNN is denoted as $\boldsymbol{X}\in {\mathbb{R}}^{N\times {F}}$, where ${F}$ represents the feature dimension for each channel.

\subsection{CGNN and IGNN}
\subsubsection{Definition of Brain Network}
A brain network is defined as an undirected graph  $\mathcal{G}=\begin{Bmatrix}\mathcal{V},\mathcal{E}\end{Bmatrix}$, where $\mathcal{V}=\begin{Bmatrix}v_1,v_2,\ldots,v_N\end{Bmatrix}$ denotes the set of nodes, corresponding to EEG channels. $N$ denotes the total number of nodes. The node feature matrix $\boldsymbol{X}$ is derived from the output of the preceding  1-D CNN. The set of edges is defined as $\mathcal{E}=\begin{Bmatrix}
(v_i,v_j) \mid i,j \in [1,N] \end{Bmatrix}$. The graph edges are denoted by a symmetric adjacency matrix $\boldsymbol{A} \in {\mathbb{R}}^{N\times N}$, where the element ${a}_{ij} > 0$ indicates the connection weight between nodes $v_i$ and $v_j$, if $(v_i,v_j) \in \mathcal{E}$. If $(v_i,v_j) \notin  \mathcal{E}$, $a_{ij}=0$. In HybGNN, two types of channel-level graphs are constructed: ${\mathcal{G}}_{C}$ and ${\mathcal{G}}_{I}$. ${\mathcal{G}}_{C}$ denotes the graph in the CGNN and ${\mathcal{G}}_{I}$ denotes the graph in the IGNN.

\subsubsection{Construction of CGNN}
 The existence of common abnormalities greatly raises the probability of depression. To uncover these common abnormal patterns, we design a common graph structure. However, the brain's complexity and the unclear mechanisms of depression make it challenging to define common graph connections directly. Therefore, instead of defining common graph connections, we make them learnable via an adjacency matrix, as inspired by \cite{dgcnn}. This matrix enables the model to explore the common graph connections from the training data automatically. Once training is complete, the matrix becomes fixed and will be applied to all input instances. The common adjacency matrix is denoted as ${\boldsymbol{A}}_{C}\in {\mathbb{R}}^{N \times N}$. ReLU function is applied to $\boldsymbol{A}_C$ to ensure that all its elements are non-negative.

\subsubsection{Construction of IGNN}  
To tackle the individual differences and reveal the individual-specific abnormal patterns for the detection of depression, we design the IGNN model, where the graph is constructed based on instance data. Let $\boldsymbol{x}_i \in \mathbb{R}^{1 \times F}$ denote the $i$-th row of $\boldsymbol{X}$. We employ a graph construction function $h(\boldsymbol{x}_i,\boldsymbol{x}_j)$, to measure the similarities between the features of two nodes, thereby forming the adjacency matrix. Each element $\boldsymbol{A}_I{(i,j)}$ of the adjacency matrix $\boldsymbol{A}_I  \in \mathbb{R}^{N \times N}$ is defined by:
\begin{gather} 
    \boldsymbol{A}_I{(i,j)}=Softmax(h(\boldsymbol{x}_{i},\boldsymbol{x}_{j}))=\frac{exp(h(\boldsymbol{x}_{i},\boldsymbol{x}_{j}))}{\textstyle\sum_{j=1}^{N}exp(h(\boldsymbol{x}_{i},\boldsymbol{x}_{j}))},  \nonumber  \\
    h(\boldsymbol{x}_{i},\boldsymbol{x}_{j})=(\boldsymbol{x}_{i}\boldsymbol{W}_{1})(\boldsymbol{x}_{j}\boldsymbol{W}_{2})^\top,
\end{gather}
where $\boldsymbol{W}_{1} \in \mathbb{R}^{F \times F_h}$ and $\boldsymbol{W}_{2} \in \mathbb{R}^{F \times F_h}$ are learnable projection matrices and $exp$ denotes exponential function. The Softmax function normalizes the similarities between nodes, ensuring that $\boldsymbol{A}_I{(i,j)}$ falls within the range of $(0,1)$.

\subsubsection{Graph Convolution Operations}
After the construction of the two graphs, we perform graph convolution operations in the two GNNs to extract richer spatial features. Generally, graph convolution  performs feature transformation on input $(\boldsymbol{X},\boldsymbol{A})$, $\boldsymbol{X} \in \mathbb{R}^{N \times F}$, $\boldsymbol{A}\in \mathbb{R}^{N \times N}$. We adopt the widely used SGC\cite{SGC} for graph convolution and modify it to incorporate multi-level graph information. The graph convolution operation in HybGNN is defined as:
\begin{align}
    {f}_{GCN}^{(L,B)}(\boldsymbol{X},\boldsymbol{A})=\displaystyle\sum_{l=0}^{L}\sigma(\hat{\boldsymbol{A}}^l \boldsymbol{X} \boldsymbol{W}^{(l,B)}),
\end{align}
where $\sigma$ denotes LeakyReLU activation function. $\hat{\boldsymbol{A}}$ is the normalized form of $\boldsymbol{A}$, defined as $\hat{\boldsymbol{A}}=\tilde{\boldsymbol{D}}^{-\frac{1}{2}} \tilde{\boldsymbol{A}} \tilde{\boldsymbol{D}}^{-\frac{1}{2}}$, where $\tilde{\boldsymbol{A}}=\boldsymbol{A}+\boldsymbol{I}$, and $\tilde{\boldsymbol{D}}$ is the degree matrix with $\tilde{\boldsymbol{D}}{(i,i)}=\sum_j \tilde{\boldsymbol{A}}{(i,j)}$. $\boldsymbol{W}^{(l,B)} \in \mathbb{R}^{{F} \times {d}}$ is a trainable linear transformation matrix at layer $l$ for branch $B$, where the superscript $B$ indicates the identifier of the branch or module. $L$ denotes number of layers.

Then, the outputs of the graph convolution in the CGNN and IGNN are given as:
\begin{align}
    &\boldsymbol{Y}_C={f}_{GCN}^{(L_1,CGNN)}(\boldsymbol{X},\boldsymbol{A}_C) \in \mathbb{R}^{N \times d}, \nonumber \\
    &\boldsymbol{Y}_I={f}_{GCN}^{(L_2,IGNN)}(\boldsymbol{X},\boldsymbol{A}_I) \in \mathbb{R}^{N \times d},
\end{align}
where $d$ represents the output feature dimension.

\subsection{Cross-Branch Hierarchical Information Extractor (CB-HIE)}

In this section, we introduce the CB-HIE module in HybGNN, which extracts valuable hierarchical information to enhance depression detection. The hierarchical information here mainly refers to the partition of nodes into regions and the interactions among regions in the region-level graph. In CB-HIE, Gated Adaptive Graph Pooling (GAGP) first partitions nodes into regions and constructs the region-level graph $\mathcal{G}_R$. Subsequently, graph convolution is used to capture the interactions among regions. Finally, to ensure that the CB-HIE output has the same dimension as the outputs of CGNN and IGNN, an adaptive graph unpooling operation is employed.

\subsubsection{Gated Adaptive Graph Pooling (GAGP)}

\begin{figure}
\centering
\includegraphics[width=0.75\linewidth]{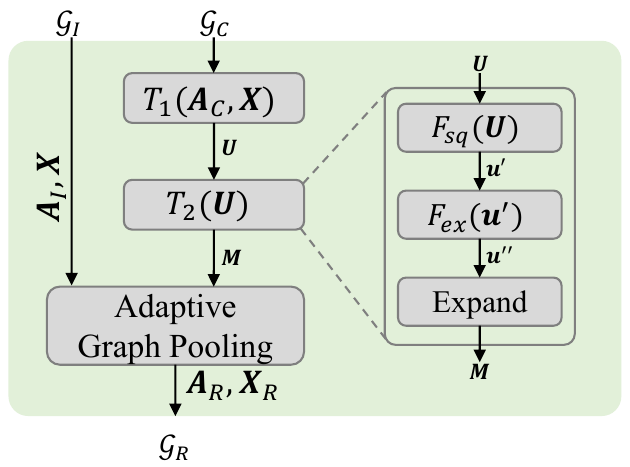} 
\caption{The framework of the proposed GAGP.}
\label{fig2}
\end{figure}


Since the partitioning of EEG channels into brain regions is not well-defined for depression detection, we propose GAGP, inspired by Diffpool\cite{diffpool}, to automatically cluster $N$ channels into $N_R$ brain regions with an assignment matrix $\boldsymbol{P} \in \mathbb{R}^{N \times N_R}$. Each row of $\boldsymbol{P}$ corresponds to an original channel-level node, and each column represents a region. The generation of $\boldsymbol{P}$ combines informative IGNN features with robust CGNN features. The process of GAGP is illustrated in Fig.~\ref{fig2} and the generation of $\boldsymbol{P}$ is formulated as:
\begin{align} \label{equ}
    & \boldsymbol{P}=Softmax((\hat{(\boldsymbol{A}_{I}}\boldsymbol{X})\circ \boldsymbol{M})\boldsymbol{Q}) \in \mathbb{R}^{N \times N_R}, \nonumber \\
    & \boldsymbol{M}=T_2(T_1(\boldsymbol{A}_C,\boldsymbol{X})) \in \mathbb{R}^{N \times F},
\end{align}
where informative $\boldsymbol{A}_I$ provides the assignment process with more individualized information. The mask $\boldsymbol{M}$ is generated using the robust CGNN features, which mitigate the impact of pseudo-connections and noise in IGNN features, enabling more task-relevant hierarchical information extraction. The matrix $\boldsymbol{Q} \in \mathbb{R}^{F \times N_R}$ is a parameter matrix to fuse the features. The symbol $\circ$ denotes the element-wise product. The function $T_1$ integrates the CGNN features, while the function $T_2$ emphasizes the most significant part of the CGNN features to suppress noise in IGNN features.

Concerning the generation of the mask $\boldsymbol{M}$, the transformation function $T_1$ is defined as:
\begin{align}
    \boldsymbol{U}=T_1(\boldsymbol{A}_C,\boldsymbol{X})=[(\boldsymbol{A}_C\boldsymbol{W}_{T_1})\mid \mid\boldsymbol{X})]  \in \mathbb{R}^{N \times 2F},
\end{align}
where $\boldsymbol{U}$ denotes the integrated graph features of $\mathcal{G}_C$, and $\boldsymbol{W}_{T_1} \in \mathbb{R}^{N \times F}$ is a learnable projection matrix which maps $\boldsymbol{A}_C$ to the same dimension as $\boldsymbol{X}$, thereby avoiding bias toward the part with the larger size. The symbol $\mid\mid$ denotes concatenation. To make $\boldsymbol{U}$ more effective in capturing and minimizing noise in $\mathcal{G}_I$ during the adaptive graph pooling, $T_2$ adopts a scheme similar to \cite{SENet}. This scheme consists of a squeeze transformation $F_{sq}$, an excitation transformation $F_{ex}$ and an expand operation. The input to $F_{sq}$ is $\boldsymbol{U} \in \mathbb{R}^{N \times 2F}$ and its output is $\boldsymbol{u}^{\prime}\in \mathbb{R}^{2F}$, where the $k$-th element of $\boldsymbol{u}^{\prime}$ is calculated by:
\begin{align}
    \boldsymbol{u}^{\prime}_{k}=F_{sq}(\boldsymbol{U})=\frac{1}{N}\displaystyle\sum_{i=1}^{N}\boldsymbol{U}(i,k),
\end{align}
where $\boldsymbol{U}(i,k)$ denotes the element of $\boldsymbol{U}$.  $\boldsymbol{u}^{\prime}$ is then passed to $F_{ex}$, producing the output $\boldsymbol{u}^{\prime\prime}$. The calculation of $F_{ex}$ is given by:
\begin{align}
    \boldsymbol{u}^{\prime\prime}=F_{ex}(\boldsymbol{u}^{\prime})=\gamma(\boldsymbol{W}_{ex_2}\delta(\boldsymbol{W}_{ex_1}\boldsymbol{u}^{\prime})) \in \mathbb{R}^{F},
\end{align}
where $\boldsymbol{W}_{ex_1} \in \mathbb{R}^{{\left \lfloor\frac{F}{r} \right \rfloor} \times 2F}$ and $\boldsymbol{W}_{ex_2} \in \mathbb{R}^{{F \times {\left \lfloor\frac{F}{r} \right \rfloor}}}$ are two learnable matrices. Here, $r$ is a hyperparameter that controls model complexity. $\delta$ denotes the ReLU function and $\gamma$ denotes the Sigmoid function. Finally, ${\boldsymbol{u}^{\prime\prime}}^\top \in \mathbb{R}^{1 \times F}$ is expanded into $\boldsymbol{M} \in \mathbb{R}^{N \times F}$ by replicating it $N$ times along the row dimension.

In the Adaptive Graph Pooling operation, the node features $\boldsymbol{X}$, the adjacency matrix $\boldsymbol{A}_I$ of $\mathcal{G}_I$ and the mask $\boldsymbol{M}$ are utilized to derive the assignment matrix $\boldsymbol{P}$ as described in \eqref{equ}. Then, $\boldsymbol{P}$ is applied to create a new coarsened adjacency matrix representing the relationships between the regions, along with an updated feature matrix for each region:
\begin{align}
    &\boldsymbol{A}_R=\boldsymbol{P}^\top\boldsymbol{A}_{I}\boldsymbol{P} \in \mathbb{R}^{N_R \times N_R}, \nonumber \\
    &\boldsymbol{X}_R=\boldsymbol{P}^\top \boldsymbol{X} \in \mathbb{R}^{N_R \times F},
\end{align}
where $\boldsymbol{A}_R$ denotes the adjacency matrix for the region-level graph and $\boldsymbol{X}_R$ denotes the features of the regions. 

\subsubsection{Graph Convolution and Adaptive Graph Unpooling}
After constructing the region-level graph $\mathcal{G}_R$, the graph convolution operation is performed to capture the interactions among the regions. The calculation is expressed as:
\begin{align}
    \boldsymbol{Y}_R=f_{GCN}^{(L_R,CB-HIE)}(\boldsymbol{X}_R,\boldsymbol{A}_R) \in \mathbb{R}^{N_R \times d},
\end{align}
where $\boldsymbol{Y}_R \in \mathbb{R}^{N_R \times d}$ represents the output of the graph convolution operation.
To integrate the region-level features with the original graph features, an unpooling operation is introduced utilizing the assignment matrix $\boldsymbol{P}$. The result is represented as:
\begin{align}
\boldsymbol{Y}_H=\boldsymbol{P}\boldsymbol{Y}_R \in \mathbb{R}^{N \times d}.
\end{align}

Subsequently, the hierarchical information $\boldsymbol{Y}_H$ is merged with the CGNN features $\boldsymbol{Y}_C$ and IGNN features $\boldsymbol{Y}_I$, maintaining the same dimensionality. The final output of the two graph models combined with CB-HIE is given by:
\begin{align}
&\boldsymbol{Y}_{all}=[\boldsymbol{Y}_C \mid \mid \boldsymbol{Y}_I \mid \mid \boldsymbol{Y}_H] \in \mathbb{R}^{N \times 3d},
\end{align}
where $\boldsymbol{Y}_{all}$ denotes the final output.

\subsection{Loss Function}
Note that the membership for each brain region should be clearly defined, i.e., one node is mapped to one region in assignment matrix $\boldsymbol{P}$. So an entropy minimization regularization is composed to each row of $\boldsymbol{P}$, encouraging the rows to be approximately one-hot vectors\cite{diffpool}. The entropy minimization loss $\mathcal{L}_{em}$ is defined as:
\begin{align}
\mathcal{L}_{em}=- \sum_{i=1}^N \sum_{j=1}^{N_R} \boldsymbol{P}{(i,j)} \log \boldsymbol{P}{(i,j)}, 
\end{align}
where $N$ is the number of nodes, $N_R$ is the number of regions and $\boldsymbol{P}{(i,j)}$ is the $i$-th row and $j$-th column of $\boldsymbol{P}$.

Consequently, our model aims to minimize both the cross-entropy loss and the entropy minimization loss of the assignment matrix $\boldsymbol{P}$. The final loss $\mathcal{L}_{final}$ is defined in the following form:
\begin{gather} 
    \mathcal{L}_{final}=\mathcal{L}_{ce}+\lambda \mathcal{L}_{em},  \nonumber  \\
    \mathcal{L}_{ce}=-\sum_{c=1}^C y_c\log \hat{y}_c,
\end{gather}
where $\mathcal{L}_{ce}$ represents the cross-entropy loss. $\lambda$ represents regularization coefficient. $C$ denotes the total number of classes, which is two in the depression detection task. $y_c$ denotes the ground truth and $\hat{y}_c$ denotes the predicted value of the model.

\section{Experiments}
\subsection{Datasets and Preprocessing}
The effectiveness of the proposed model is validated using two publicly available datasets. We mainly focus on the resting-state parts of EEG data in the two datasets.

\textbf{1)MODMA:} Multi-modal Open Dataset for Mental Disorder Analysis
(MODMA)\cite{modma} comprises EEG recordings from 24 individuals with Major Depressive Disorder (MDD) and 29 Healthy Controls (HCs). Each participant underwent recording for five minutes using a 128-channel EEG cap.

\textbf{2)HUSM:} The Hospital University Sains Malaysia (HUSM)\cite{HUSM} dataset comprises EEG recordings from 34 MDD individuals and 30 HCs. Each participant underwent recording for five minutes using a 19-channel EEG cap with eyes closed and open respectively.

We first perform preprocessing following the previous studies\cite{2021deprnet}. For the MODMA dataset, we only use the 19 channels in the 10-20 system to be consistent with HUSM dataset and to reduce computational complexity. The EEG recordings are segmented into non-overlapping four-seconds segments, resulting in 1,728 segments from individuals with MDD and 2,088 segments from HCs. For the HUSM dataset, we apply the same segmentation process to the EEG recordings, combining both eyes-closed and eyes-open data. After removing subjects with missing data, we obtain 3,726 MDD segments and 3,588 HC segments.

\subsection{Implementation Details}
We follow previous studies \cite{hubin2022gicn,62SDGCN,24SGPSL} and perform a ten-fold cross-validation on each dataset to obtain reliable evaluation results. Furthermore, to prevent information leakage, the data from a subject is used exclusively in either the training set or the test set. The code is publicly available on \url{https://github.com/wyyadams/HybGNN\_DeprDetection}.

\begin{figure}[ht]
    \centering
    \begin{minipage}[t]{0.24\textwidth}
        \centering
        \includegraphics[width=\textwidth]{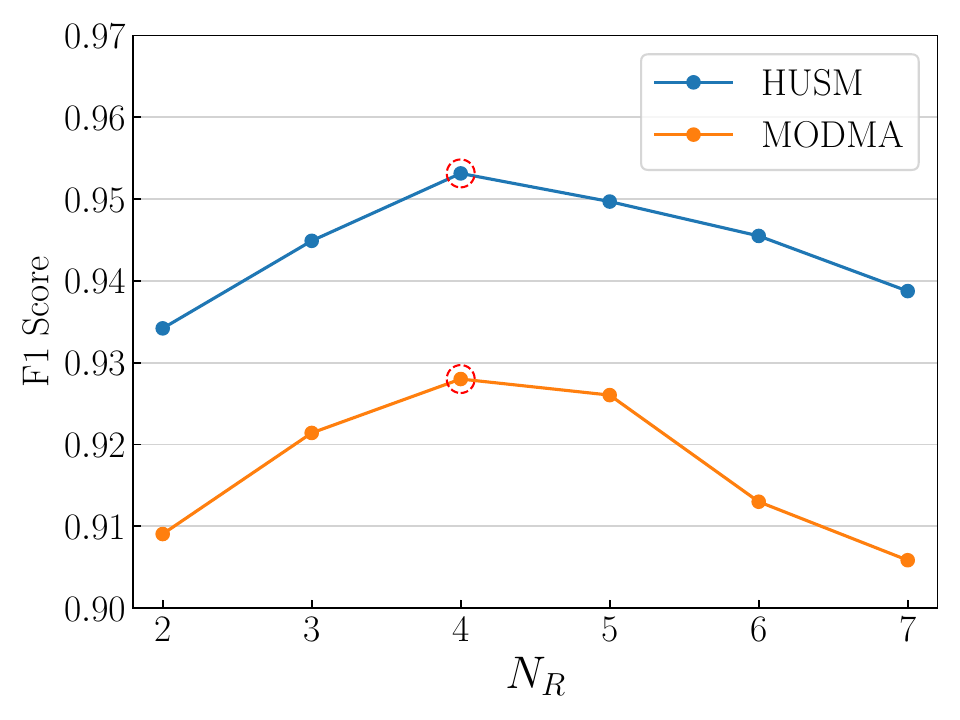} 
    \end{minipage}
    \hfill
    \begin{minipage}[t]{0.24\textwidth}
        \centering
        \includegraphics[width=\textwidth]{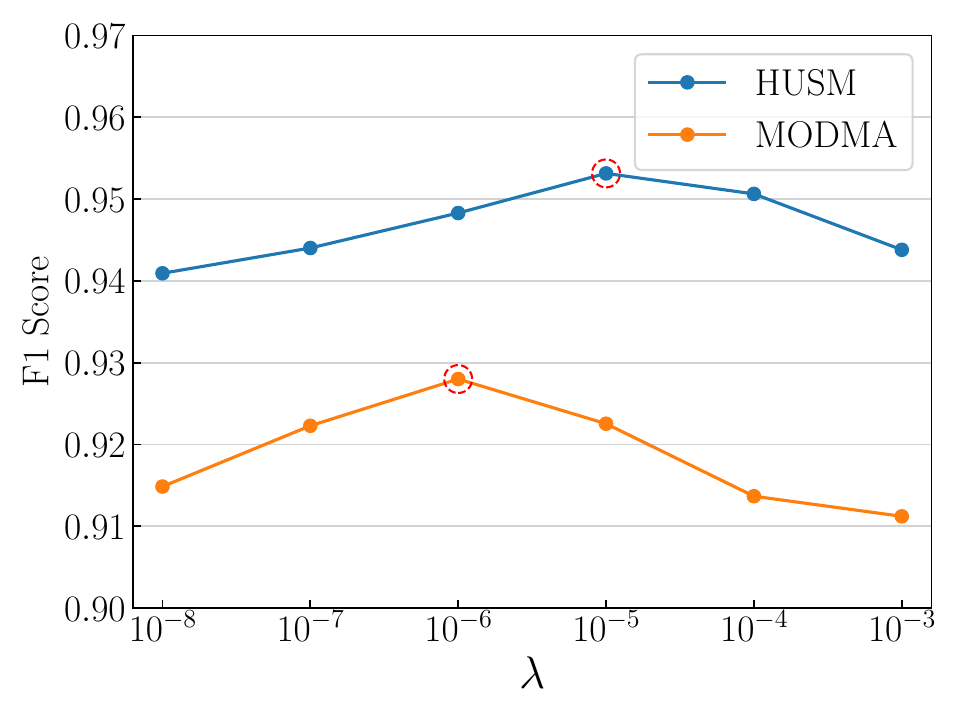}
    \end{minipage}
    \caption{Hyperparameter optimization for the number of regions $N_R$ and regularization coefficient $\lambda$.}
    \label{fig:both}
\end{figure}

To thoroughly assess depression detection performance, accuracy (ACC), recall (REC), precision (PRE), and F1 score (F1) are used as evaluation metrics in experiments. We implement HybGNN using PyTorch libraries\cite{pytorch} on an NVIDIA GeForce RTX 4060. The detailed configuration of hyperparameters is presented in Table~\ref{tab:config}. The results of hyperparameter optimization for $N_R$ and $\lambda$ are illustrated in Fig.~\ref{fig:both}.

\begin{table}
\centering
\caption{Configuration of hyperparameters}
\label{tab:config}
\resizebox{0.48\textwidth}{!}{
\begin{tabular}{|c|c|cc|}
\hline
\multicolumn{1}{|c|}{\multirow{2}{*}{\textbf{HParam$^{\mathrm{a}}$}}} &
  \multicolumn{1}{c|}{\multirow{2}{*}{\textbf{Description}}} &
  \multicolumn{2}{c|}{\textbf{Value}} \\ \cline{3-4} 
\multicolumn{1}{|c|}{} &
  \multicolumn{1}{c|}{} &
  \multicolumn{1}{c|}{\textit{\textbf{MODMA}}} &
  \textit{\textbf{HUSM}} \\ \hline
Epoch  & Total number of epochs during training                  & \multicolumn{1}{c|}{100}       & 100       \\ \hline
Lr     & Learning rate                                           & \multicolumn{1}{c|}{0.03}      & 0.001     \\ \hline
Bs     & Batch size                                              & \multicolumn{1}{c|}{64}        & 128       \\ \hline
$L_1$     & Number of GCN layers in CGNN                            & \multicolumn{1}{c|}{2}         & 2         \\ \hline
$L_2$     & Number of GCN layers in IGNN                            & \multicolumn{1}{c|}{2}         & 2         \\ \hline
$N_R$     & Number of regions in CB-HIE                           & \multicolumn{1}{c|}{4}         & 4         \\ \hline
$r$      & The HParam to control model complexity in CB-HIE & \multicolumn{1}{c|}{16}        & 16        \\ \hline
$L_R$     & Number of GCN layers in CB-HIE                         & \multicolumn{1}{c|}{1}         & 1         \\ \hline
$\lambda$ & Regularization coefficient                              & \multicolumn{1}{c|}{$10^{-6}$} & $10^{-5}$ \\ \hline
\multicolumn{4}{l}{$^{\mathrm{a}}$HParam is the abbreviation for hyperparameter.}
\end{tabular}
}
\end{table}

\subsection{Compared Method}
To assess the effectiveness of the proposed HybGNN, we perform a comprehensive comparison with various methods utilized in depression detection or widely used in EEG-related tasks: DeprNet\cite{2021deprnet}, DGCNN\cite{dgcnn}, GICN\cite{hubin2022gicn}, SDGCN\cite{62SDGCN}, GraphSleepNet\cite{graphsleepnet}, LGGNet\cite{ding2023LGGNet}, AHGNN\cite{xue2022ahgnn}. A summary of their key characteristics is provided in Table~\ref{tab:Comparison}. Notably, these methods differ in their architectural foundations and application domains. DeprNet employs a CNN-based architecture, whereas the remaining methods utilize GNNs. DGCNN and AHGNN are developed for EEG emotion recognition. LGGNet is designed for brain-computer interface. GraphSleepNet is tailored for sleep stage classification. The other methods, including DeprNet, GICN, and SDGCN, are specifically designed for EEG-based depression detection. In addition, LGGNet, AHGNN and the proposed HybGNN adopt distinct processes for constructing the region-level graph as described in section \ref{gnnec}. For the aforementioned GNN-based methods, all graph models employ the same 1-D CNN (proposed in section \ref{2.2}) as the temporal feature extractor. To ensure a convincing comparison with the proposed method, we directly execute (or reproduce) the codes of the compared methods on both datasets.

\section{Results and Discussion}
\subsection{Comparison of Depression Detection Performance}
Table~\ref{tab:Comparison} summarizes the detection performance comparison on the MODMA and HUSM datasets. The proposed HybGNN consistently outperforms other methods across most metrics, demonstrating its effectiveness in depression detection. Specifically, HybGNN achieves superior detection results. On the MODMA dataset, it gains a 0.015 higher F1 score compared to LGGNet, which ranks second in F1 score. On the HUSM dataset, it achieves a 0.024 higher F1 score compared to AHGNN.

Additionally, several other conclusions can be drawn from the comparison. First, the comparison between the CNN-based DeprNet and the graph-based methods shows that GNN-based methods outperform CNN-based methods for depression detection, which is consistent with findings from other EEG-related classification tasks. Second, a comparison of GraphSleepNet's results on the HUSM dataset with those of DGCNN, GICN, and SDGCN reveals that individualized graph connections do not always outperform common ones, in contrast to findings from other EEG-related tasks\cite{graphsleepnet,IAG}. This discrepancy is likely attributed to the high variability among depression patients, necessitating extensive data for optimal model training. However, EEG data for depression detection remains limited compared to other tasks, such as EEG emotion recognition. Consequently, purely adaptive individualized graph connections are susceptible to noise and overfitting, highlighting the importance of combining them with common graph connections. Finally, the superior performance of LGGNet and AHGNN over DGCNN, GICN, and SDGCN across most evaluation metrics confirms the potential of incorporating region-level graphs and hierarchical information to enhance depression detection. Furthermore, the proposed HybGNN surpasses LGGNet and AHGNN, demonstrating that the refined construction of region-level graphs and the integration of individualized information further amplify the effectiveness of hierarchical information in improving detection accuracy.

\begin{table*}
\centering
\caption{Performance comparison of detection methods}
\label{tab:Comparison}
\begin{tabular}{|c|cc|c|cccc|cccc|}
\hline
\multirow{2}{*}{\textbf{Methods}} &
  \multicolumn{2}{c|}{\textbf{Graph Connections}} &
  \textbf{Region-Level} &
  \multicolumn{4}{c|}{\textbf{MODMA}} &
  \multicolumn{4}{c|}{\textbf{HUSM}} \\ \cline{2-3} \cline{5-12} 
 &
  \multicolumn{1}{c|}{\textit{\textbf{Common}}} &
  \textit{\textbf{Individualized}} &
  \textbf{Graph} &
  \multicolumn{1}{c|}{\textit{\textbf{ACC(\%)}}} &
  \multicolumn{1}{c|}{\textit{\textbf{REC(\%)}}} &
  \multicolumn{1}{c|}{\textit{\textbf{PRE(\%)}}} &
  \textit{\textbf{F1}} &
  \multicolumn{1}{c|}{\textit{\textbf{ACC(\%)}}} &
  \multicolumn{1}{c|}{\textit{\textbf{REC(\%)}}} &
  \multicolumn{1}{c|}{\textit{\textbf{PRE(\%)}}} &
  \textit{\textbf{F1}} \\ \hline
DeprNet &
  \multicolumn{1}{c|}{} &
   &
   &
  \multicolumn{1}{c|}{85.36} &
  \multicolumn{1}{c|}{79.89} &
  \multicolumn{1}{c|}{93.60} &
  0.852 &
  \multicolumn{1}{c|}{87.68} &
  \multicolumn{1}{c|}{84.68} &
  \multicolumn{1}{c|}{91.20} &
  0.874 \\ \hline
DGCNN &
  \multicolumn{1}{c|}{$\checkmark$} &
   &
   &
  \multicolumn{1}{c|}{89.88} &
  \multicolumn{1}{c|}{87.55} &
  \multicolumn{1}{c|}{91.66} &
  0.890 &
  \multicolumn{1}{c|}{91.06} &
  \multicolumn{1}{c|}{92.88} &
  \multicolumn{1}{c|}{91.28} &
  0.915 \\ \hline
GICN &
  \multicolumn{1}{c|}{$\checkmark$} &
   &
   &
  \multicolumn{1}{c|}{89.00} &
  \multicolumn{1}{c|}{84.71} &
  \multicolumn{1}{c|}{95.83} &
  0.893 &
  \multicolumn{1}{c|}{92.05} &
  \multicolumn{1}{c|}{91.27} &
  \multicolumn{1}{c|}{90.80} &
  0.916 \\ \hline
SDGCN &
  \multicolumn{1}{c|}{$\checkmark$} &
   &
   &
  \multicolumn{1}{c|}{90.31} &
  \multicolumn{1}{c|}{88.58} &
  \multicolumn{1}{c|}{94.46} &
  0.911 &
  \multicolumn{1}{c|}{90.73} &
  \multicolumn{1}{c|}{92.42} &
  \multicolumn{1}{c|}{90.19} &
  0.907 \\ \hline
GraphSleepNet &
  \multicolumn{1}{c|}{} &
  $\checkmark$ &
   &
  \multicolumn{1}{c|}{90.84} &
  \multicolumn{1}{c|}{86.76} &
  \multicolumn{1}{c|}{\textbf{96.86}} &
  0.912 &
  \multicolumn{1}{c|}{90.90} &
  \multicolumn{1}{c|}{90.91} &
  \multicolumn{1}{c|}{89.63} &
  0.900 \\ \hline
LGGNet &
  \multicolumn{1}{c|}{$\checkmark$} &
   &
  $\checkmark$ &
  \multicolumn{1}{c|}{90.47} &
  \multicolumn{1}{c|}{88.26} &
  \multicolumn{1}{c|}{95.39} &
  0.913 &
  \multicolumn{1}{c|}{92.15} &
  \multicolumn{1}{c|}{92.70} &
  \multicolumn{1}{c|}{92.13} &
  0.920 \\ \hline
AHGNN &
  \multicolumn{1}{c|}{$\checkmark$} &
   &
  $\checkmark$ &
  \multicolumn{1}{c|}{90.57} &
  \multicolumn{1}{c|}{87.84} &
  \multicolumn{1}{c|}{93.75} &
  0.904 &
  \multicolumn{1}{c|}{92.84} &
  \multicolumn{1}{c|}{93.45} &
  \multicolumn{1}{c|}{92.99} &
  0.929 \\ \hline
Ours &
  \multicolumn{1}{c|}{$\checkmark$} &
  $\checkmark$ &
  $\checkmark$ &
  \multicolumn{1}{c|}{\textbf{92.42}} &
  \multicolumn{1}{c|}{\textbf{91.31}} &
  \multicolumn{1}{c|}{95.30} &
  \textbf{0.928} &
  \multicolumn{1}{c|}{\textbf{95.37}} &
  \multicolumn{1}{c|}{\textbf{96.80}} &
  \multicolumn{1}{c|}{\textbf{94.22}} &
  \textbf{0.953} \\ \hline
\end{tabular}
\end{table*}

\subsection{Ablation Study}

\begin{table*}
\centering
\caption{The results of the ablation study}
\label{tab:ablation}
\begin{tabular}{|c|cccc|cccc|}
\hline
\multirow{2}{*}{\textbf{Methods}} &
  \multicolumn{4}{c|}{\textbf{MODMA}} &
  \multicolumn{4}{c|}{\textbf{HUSM}} \\ \cline{2-9} 
 &
  \multicolumn{1}{c|}{\textit{\textbf{ACC(\%)}}} &
  \multicolumn{1}{c|}{\textit{\textbf{REC(\%)}}} &
  \multicolumn{1}{c|}{\textit{\textbf{PRE(\%)}}} &
  \textit{\textbf{F1}} &
  \multicolumn{1}{c|}{\textit{\textbf{ACC(\%)}}} &
  \multicolumn{1}{c|}{\textit{\textbf{REC(\%)}}} &
  \multicolumn{1}{c|}{\textit{\textbf{PRE(\%)}}} &
  \textit{\textbf{F1}} \\ \hline
\textit{Variant a} &
  \multicolumn{1}{c|}{89.80} &
  \multicolumn{1}{c|}{87.66} &
  \multicolumn{1}{c|}{92.49} &
  0.895 &
  \multicolumn{1}{c|}{91.56} &
  \multicolumn{1}{c|}{92.49} &
  \multicolumn{1}{c|}{91.71} &
  0.918 \\ \hline
\textit{Variant b} &
  \multicolumn{1}{c|}{89.31} &
  \multicolumn{1}{c|}{85.89} &
  \multicolumn{1}{c|}{96.74} &
  0.904 &
  \multicolumn{1}{c|}{92.05} &
  \multicolumn{1}{c|}{92.73} &
  \multicolumn{1}{c|}{92.90} &
  0.926 \\ \hline
\textit{Variant c} &
  \multicolumn{1}{c|}{90.13} &
  \multicolumn{1}{c|}{86.67} &
  \multicolumn{1}{c|}{96.34} &
  0.909 &
  \multicolumn{1}{c|}{92.21} &
  \multicolumn{1}{c|}{90.77} &
  \multicolumn{1}{c|}{\textbf{96.78}} &
  0.934 \\ \hline
\textit{Variant d} &
  \multicolumn{1}{c|}{91.44} &
  \multicolumn{1}{c|}{88.58} &
  \multicolumn{1}{c|}{94.06} &
  0.909 &
  \multicolumn{1}{c|}{94.15} &
  \multicolumn{1}{c|}{95.04} &
  \multicolumn{1}{c|}{92.48} &
  0.932 \\ \hline
\textit{Variant e} &
  \multicolumn{1}{c|}{91.60} &
  \multicolumn{1}{c|}{87.53} &
  \multicolumn{1}{c|}{\textbf{96.84}} &
  0.915 &
  \multicolumn{1}{c|}{94.65} &
  \multicolumn{1}{c|}{95.85} &
  \multicolumn{1}{c|}{92.75} &
  0.941 \\ \hline
\textit{Variant f} &
  \multicolumn{1}{c|}{91.43} &
  \multicolumn{1}{c|}{88.85} &
  \multicolumn{1}{c|}{96.56} &
  0.922 &
  \multicolumn{1}{c|}{94.47} &
  \multicolumn{1}{c|}{94.71} &
  \multicolumn{1}{c|}{94.67} &
  0.944 \\ \hline
Ours &
  \multicolumn{1}{c|}{\textbf{92.42}} &
  \multicolumn{1}{c|}{\textbf{91.31}} &
  \multicolumn{1}{c|}{95.30} &
  \textbf{0.928} &
  \multicolumn{1}{c|}{\textbf{95.37}} &
  \multicolumn{1}{c|}{\textbf{96.80}} &
  \multicolumn{1}{c|}{94.22} &
  \textbf{0.953} \\ \hline
\end{tabular}
\end{table*}

To further validate the key components in our model, we perform an ablation study. In the experiment, we initially consider the CGNN as the baseline framework. Subsequently, we progressively integrate additional components, including IGNN and CB-HIE to construct the proposed model. Additionally, we explore the optimal integration of the Hierarchical Information Extractor (HIE) module. 

Specifically, \textit{Variant a, b, c} are generated to verify the effectiveness of CGNN, IGNN and CB-HIE respectively, with \textit{Variant a} serving as the baseline.  \textit{Variant d, e, f} are generated to explore the optimal integration of the Hierarchical Information Extractor (HIE) and compare with our HybGNN. The following outlines the specific details for each variant:

\begin{itemize}
    \item \textit{Variant a}: CGNN.
    \item \textit{Variant b}: IGNN.
    \item \textit{Variant c}: CGNN + IGNN.
    \item \textit{Variant d}: CGNN + IGNN + C-HIE.
    \item \textit{Variant e}: CGNN + IGNN + I-HIE.
    \item \textit{Variant f}: CGNN + IGNN + CB-HIE-I.
    \item Ours: CGNN + IGNN + CB-HIE.
\end{itemize}

Notably, the difference among \textit{Variant d, e, f} and ours lies in \eqref{equ}. Compared with our HybGNN, C-HIE replaces $\hat{\boldsymbol{A}_{I}}$ with $\hat{\boldsymbol{A}_{C}}$ and omits $\boldsymbol{M}$. I-HIE omits $\boldsymbol{M}$ directly. In CB-HIE-I, the $\hat{\boldsymbol{A}_{I}}$ is replaced with $\hat{\boldsymbol{A}_{C}}$ and $\boldsymbol{M}$ is generated by $T_2(T_1(\boldsymbol{A}_I,\boldsymbol{X}))$.

The results are presented in Table~\ref{tab:ablation}. First, by comparing \textit{Variant a, b} with \textit{Variant c}, we notice a decrease in depression detection accuracy on both datasets without CGNN or IGNN. This leads us to conclude that CGNN and IGNN together improve the accuracy of depression detection. Second, comparing the \textit{Variant c} with our model reveals that CB-HIE module effectively aids in extracting hierarchical information, thereby improving depression detection accuracy. Additionally, by comparing the \textit{Variant d, e, f} with our model, we infer that constructing individualized region-level graph based on IGNN features, with the CGNN features serving as the input of gating mechanism, yields the best performance. In summary, the experiment demonstrates the effectiveness of each key component in HybGNN.

\subsection{Visualization of Degree Centrality}
\begin{figure}
\centering
\includegraphics[width=\linewidth]{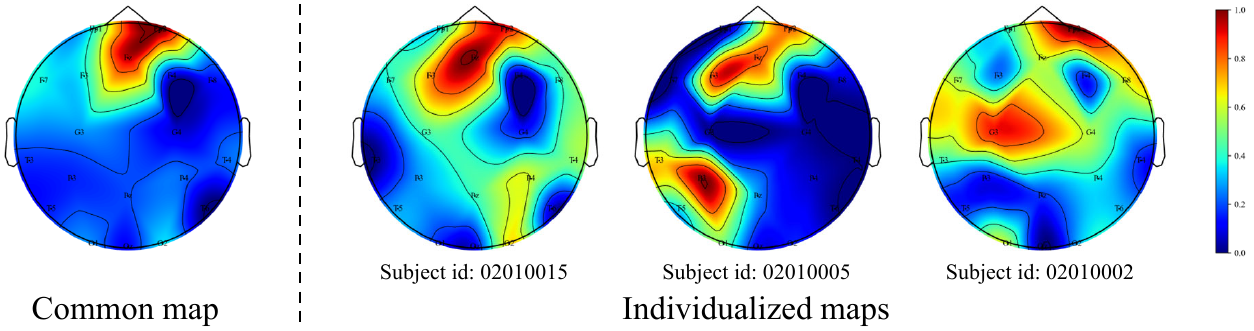} 
\caption{The visualization of degree centrality of EEG channels on MODMA.}
\label{fig4}
\end{figure}

To better illustrate the common and individualized graph connections captured by HybGNN, we visualize the degree centrality of each EEG channel on the MODMA dataset. Degree centrality is a measure of a node's importance in a graph, reflecting the strength of its connections with other nodes \cite{zhang2007ontology}. EEG channels with higher centrality values are more crucial for depression detection. The degree centrality $\boldsymbol{C}_i$ of the $i$-th EEG channel is calculated as:
\begin{align}
\boldsymbol{C}_i=\textstyle\sum_{n=1}^{N}\boldsymbol{A}(i,n)+\textstyle\sum_{m=1}^{N}\boldsymbol{A}(m,i)-2\boldsymbol{A}(i,i),
\end{align}
where $\boldsymbol{A}(i,n)$ denotes the element in the $i$-th row and $n$-th column of the adjacency matrix $\boldsymbol{A} \in \mathbb{R}^{N \times N}$. In Fig.~\ref{fig4}, the common map is generated using $\boldsymbol{A}_C$ and highlights that channels in prefrontal lobe are crucial for capturing common abnormalities among depression patients. This finding is consistent with previous research\cite{sun2019graph}, which indicates that depression patients commonly exhibit abnormalities in prefrontal lobe. The individualized maps, generated by $\boldsymbol{A}_I$ from several depressive subjects, reveal that in addition to the prefrontal lobe, channels in other brain areas, such as parietal lobe and central lobe, also contribute to the depression detection for certain individuals. In summary, the visualization confirms that HybGNN is capable of capturing common and individualized abnormalities in depressive patients.

\section{Conclusion}
In this paper, we introduce a hybrid graph neural network (HybGNN) for depression detection using EEG data. Our HybGNN framework integrates both a common graph neural network (CGNN) branch and an individualized graph neural network (IGNN) branch to identify depression-related patterns while accommodating individual variations. Additionally, the CB-HIE module combines the features from CGNN and IGNN to extract hierarchical information, enhancing depression detection. Extensive experiments and analysis on two datasets confirm the effectiveness of HybGNN. Furthermore, HybGNN shows great potential for improving depression detection in other modalities, such as fMRI, and for discovering new biomarkers for depression detection, which we plan to explore in future work.

\bibliographystyle{IEEEtran}
\bibliography{ref.bib}

\end{document}